\documentclass[twocolumn,showpacs,superscriptaddress,preprintnumbers,amsmath,amssymb,prc]{revtex4-2}

\usepackage{graphicx}
\usepackage{dcolumn}
\usepackage{bm}
\usepackage{float}
\usepackage{color}
\usepackage{CJK}
\usepackage{ulem}
\usepackage{array} 
\usepackage{makecell}
\usepackage{wrapfig}
\usepackage{lipsum} 
\usepackage{textcomp}

\usepackage[colorlinks,linkcolor=blue,urlcolor=blue,citecolor=blue]{hyperref}

\usepackage{tikz,xcolor,hyperref}

\definecolor{lime}{HTML}{A6CE39}
\DeclareRobustCommand{\orcidicon}{
	\begin{tikzpicture}
	\draw[lime, fill=lime] (0,0) 
	circle [radius=0.16] 
	node[white] {{\fontfamily{qag}\selectfont \tiny ID}};
	\draw[white, fill=white] (-0.0625,0.095) 
	circle [radius=0.007];
	\end{tikzpicture}
	\hspace{-2mm}
}
\foreach \x in {A, ..., Z}{%
	\expandafter\xdef\csname orcid\x\endcsname{\noexpand\href{https://orcid.org/\csname orcidauthor\x\endcsname}{\noexpand\orcidicon}}
}

\begin{document}
\begin{CJK*}{UTF8}{gbsn}

\title{Momentum correlation of light nuclei in Au + Au collisions
at \(\sqrt{s_{NN}}\) = 2.0 $\sim$ 7.7 GeV }

\author{Feng-Hua Qiao(乔枫华)}
\affiliation{School of Physical Science and Technology, ShanghaiTech University, Shanghai 201210, China}
\affiliation{Shanghai Institute of Applied Physics, Chinese Academy of Sciences, Shanghai 201800, China}
\affiliation{University of the Chinese Academy of Sciences, Beijing 100080, China}

\author{X. G. Deng(邓先概)\orcidB{}} 
\email[Corresponding author: ]{ xiangai\_deng@fudan.edu.cn}
\affiliation{Key Laboratory of Nuclear Physics and Ion-beam Application (MOE), Institute of Modern Physics, Fudan University, Shanghai 200433, China}
\affiliation{Shanghai Research Center for Theoretical Nuclear Physics， NSFC and Fudan University, Shanghai 200438, China}

\author{Y. G. Ma(马余刚)\orcidC{}}
\email[Corresponding author: ]{mayugang@fudan.edu.cn}
\affiliation{Key Laboratory of Nuclear Physics and Ion-beam Application (MOE), Institute of Modern Physics, Fudan University, Shanghai 200433, China}
\affiliation{Shanghai Research Center for Theoretical Nuclear Physics， NSFC and Fudan University, Shanghai 200438, China}

\date{\today}

\begin{abstract}

Within the Ultra-relativistic Quantum Molecular Dynamics model (UrQMD) coupled with nucleon coalescence model and Mini-Spanning-Tree model, the yields of light nuclei have been stimulated in Au + Au collisions over an energy range of \(\sqrt{s_{NN}}=2.0\sim7.7\ \rm{GeV}\) and the momentum correlation functions of light nuclei pairs have been calculated by both the Lednick\'{y}-Lyuboshitz and the Correlation After Burner methods. We compare the yields of light nuclei and their momentum correlation functions at midrapidity in this energy region with experimental data. It is found that there are differences between the results of the two models, and the coalescence method seems  less valid at low collision energy. Furthermore, both the peak values of proton-proton correlation functions and the transition point of elliptic flows from out-of-plane to in-plane emission show a turning point around 3-4 GeV, which suggests that there is a relation between momentum correlation function and collective flow of particles.

\end{abstract}

\pacs{}

\maketitle
\end{CJK*}

\section{Introduction}
\label{introduction}

Heavy-ion collisions (HICs) seek to investigate the characteristics of nuclear matter under different hot and dense conditions, as well as to understand the underlying principles of matter formation. In the extreme environment created by HICs, the production of light clusters with low binding energy has attracted a wide range of research interest, which offers opportunities to examine the local baryon density~\cite{Csernai,Scheibl,Bellini}, explore the properties of quantum chromo-dynamics (QCD) matter~\cite{Andronic:2010,SunKJ:2018,Braun-Munzinger} and search for the critical point~\cite{Steinheimer,Herold,NT_LuoXF,NT_ChenQ,NT_SunKJ,NT_LiF,NT_WuYF}. The theoretical frameworks for light nuclei production can be broadly classified into two primary categories: one is the thermodynamic model which considers the light nuclei being formed in a thermally and chemically equilibrated  source~\cite{Bebie,Braun-Munzinger:1995}; the other is the coalescence or cluster model which assumes that light nuclei are formed through proton-neutron pairing at the kinetic freeze-out stage of collision or evolution process~\cite{Awes,Mattiello:1996,Chen:2003prc,YanTZ}. Integrated with various HICs models, the coalescence or cluster algorithm offers a reasonably accurate description of light nuclei yield across a wide range of collision  energies~\cite{Zhao:2021,SunKJ:2021,Reichert:2022,Cheng:2021,Coci:2023,WangTT}. However, the mechanisms behind light nuclei production remain a topic of discussion. In the intermediate energy examined in this paper, specifically for the collision energy in center-of-mass frame  ($\sqrt{s_{NN}}=2.0\sim7.7\ \rm{GeV}$), the yield of light nuclei such as deuteron or triton is markedly elevated compared to other energy regions. These enhancement results lead to a higher $d/p$ or $t/p$ ratio, necessitating strict consideration of baryon number conservation. The models that describe light nuclei formation based on proton-neutron pair information, may yield diverse outcomes and warrant further discussion.

In HICs, the Hanbury Brown and Twiss (HBT) method for two-particle intensity interferometry~\cite{RH1,RH2} has been extensively employed to probe for the shape of the thermal source formed through the collision and study the reaction dynamics, owing to its sensitivity to particle emission sources~\cite{Koonin:1977,Lisa:2005,Wiedemann:1999,Wei04,Ma06,He:2020,Ghetti:2003}. Given the extremely short timescale of collisions, direct observation of the reaction zone is infeasible. Nonetheless, by extracting  particle distribution information in the later stage of the reaction and measuring the HBT correlation function, the information of nuclei such as the size of the reaction source, particle interactions, and emission time can be obtained. Experimentally, HBT results for pions and baryons have been extensively measured and analyzed across a broad spectrum of collision energies~\cite{STAR:2015,STAR:2018,ALICE:2020}. Furthermore, correlations between protons, as well as between protons and some light fragments, have been employed to probe the properties of the emitting source~\cite{Cao:2012,STAR:2014prl}. Theoretically, the dependence of correlation functions on several parameters, encompassing beam energy, collision characteristics, total momentum of the nucleon pair, the equation of state for nuclear matter, and the size of the reaction system, have also been investigated in various Refs.~\cite{Wang:2018,Wang:2019,Wang:2022,Zhang:2017}. Moreover, studies have been conducted on correlations of various particle pairs, as documented in Refs.~\cite{Lednicky:1995,Xi:2019,WangY:2021,WangTT,Wang:2024}. These studies suggest that two-particle correlation functions prove instrumental in advancing understanding of heavy-ion collisions, including the formation mechanisms of light nuclei. Recently, the second stage of the Beam Energy Scan (BES-II) program at the Relativistic Heavy Ion Collider (RHIC) was initiated, the STAR experiment conducted a fixed-target program to scan collision energies of \(\sqrt{s_{NN}}=3.0\sim7.7\ \rm{GeV}\) and the relevant experimental data including the yield and correlation of light nuclei, are being analyzed and published. Concurrently, theoretical studies have been undertaken in this energy region, aiming to explore the physical mechanisms underlying the heavy-ion collisions~\cite{Chatterjee:2020,Li:2022,Buyukcizmeci:2023}, including the investigation of particle yields using diverse models and examining their collective flow \cite{flow0,flow1,flow2,flow3,Shi,Xiao,WangH,Lan,WangM}.

To investigate the mechanisms of light nuclei production and their HBT correlation characteristics, we employ the Ultra-relativistic Quantum Molecular Dynamics (UrQMD) microscopic transport model, which incorporates a hard density-dependent equation-of-state, to simulate Au + Au collisions within an energy range of \(\sqrt{s_{NN}}=2.0\sim7.7\ \rm{GeV}\). Light nuclei production is facilitated by coupling the coalescence and the Mini-Spanning-Tree scenarios. We present the yield and momentum correlation functions of light nuclei at midrapidity across various collision centralities within this energy range, and compare them with experimental data. Notably, we discerned differences between the two scenarios, offering detailed explanations and delving into the mechanisms of each. Additionally, by contrasting the light nuclei correlation functions with the elliptic flow, we establish a linkage between particle momentum correlation functions and collective flows.

This work is organised as follows: firstly, we introduce the UrQMD model and the light nuclei formation methods we used in Sec.~\ref{UrQMD and Light nuclei}; then the momentum correlation function calculation methods: the Lednick\'{y}-Lyuboshitz and the Correlation after burner model are described in Sec.~\ref{HBT model}. In Sec.~\ref{resultsA1}, the results of rapidity distribution of light nuclei by different models are compared to the RHIC-STAR and the HADES experimental results. And Sec.~\ref{resultsB1} gives the comparison of proton-proton and deuteron-deuteron correlations with the STAR data at \(\sqrt{s_{NN}}=2.4,  3.2\ \rm{GeV}\), the differences by the algorithms between the coalescence and Mini-Spanning-Tree models are discussed. Finally, we show the calculation results of light nuclei correlation in the energy range of \(2.0\sim7.7\) GeV and connect particle correlation functions to collective flows in Sec.~\ref{resultsB2}.

\section{Methods}
\label{methods}

\subsection{UrQMD model and cluster formation}
\label{UrQMD and Light nuclei}

We employ the UrQMD model~\cite{Bass:1998,Bleicher:1999}, which is extensively used in heavy-ion collisions~\cite{Li:2022,Bleicher:2022,WangYJ:2020}, to simulate Au + Au collisions at collision energy \(\sqrt{s_{NN}}=2.0\sim7.7\ \rm{GeV}\) and output the initial distribution of particles. The UrQMD model operates in several modes; one is the default mode, which neglects potentials between nucleons during the hadronic scattering phase. For enhanced accuracy, we use the mode which contains density-dependent potentials, including the hard Skyrme equation-of-state (EOS) below \(\sqrt{s_{NN}}=3.3\ \rm{GeV}\). The Skyrme potential for two- and three-body interactions can be expressed as \(U=\alpha(\frac{\rho}{\rho_0})+\beta(\frac{\rho}{\rho_0})^{\gamma}\), where \(\rho\) denotes the baryonic interaction density. The stiffness of EOS is determined by the values of \(\alpha,\ \beta,\ \gamma\). As energy increases, collision becomes dominant and Skyrme potential can be ignored which is switched off  above \(\sqrt{s_{NN}}=3.3\ \rm{GeV}\) in UrQMD model.

The coalescence model works for the formation of light nuclei at the kinetic freeze-out stage of a collision, when hadronic elastic and inelastic scattering cease. It has been used extensively for describing the production of light clusters in heavy-ion collisions at both intermediate and high energies~\cite{Csernai,Nagle:1996,Oh:2007}. In the coalescence model we used, the probability of producing an $M$-nucleon cluster is determined by its Wigner phase-space density and the nucleon phase-space distribution at the freeze-out stage~\cite{Chen:2003prc}. The multiplicity of an $M$-hadron cluster in a heavy-ion collision is given by:
\begin{equation}
\begin{split}
 N_M = G \int d\bm{r}_{i_1} d\bm{q}_{i_1}...d\bm{r}_{i_{M-1}} d\bm{q}_{i_{M-1}} \\
 \prod_{i=1}^{i=M} \rho_i^\omega (\bm{r}_{i_1}, \bm{q}_{i_1}...\bm{r}_{i_{M-1}}, \bm{q}_{i_{M-1}}),
\end{split}                              
\end{equation}
where $G$ is the spin-isospin statistical factor and represents 3/8 for deuteron and 1/3 for triton or $^3$He. By taking the hadron wave functions to be a spherical harmonic oscillator, we can get the deuteron Wigner phase-space density~\cite{Nagle:1996}:
\begin{equation}
\begin{split}
\rho_d^{\omega}(\bm{\rho},\bm{p}_{\rho}) = 8{\rm exp}(-\frac{\bm{\rho}^2}{b^2}-{\bm{p}_{\rho}}^2b^2),
\end{split}                              
\label{deuteron-wigner function}
\end{equation}
where $\bm{\rho} = \frac{1}{\sqrt{2}}(\bm{r_1}-\bm{r_2})$ and $\bm{p}_{\rho} = \frac{1}{\sqrt{2}}(\bm{p_1}-\bm{p_2})$ are the relative coordinate and momentum of proton-neutron after Jacobi matrix transform to the center-of-mass coordinate, respectively. Following the Ref.~\cite{Chen:2003prc}, the triton (\(^3\)He) Wigner function is:
\begin{equation}
\begin{aligned}[b]
 &\rho_{t({^3}He)}^{\omega}(\bm{\rho},\bm{\lambda}, \bm{p}_{\rho}, \bm{p}_{\lambda}) \\
 &=64 \exp\left(-\frac{\bm{\rho}^2+\bm{\lambda}^2}{b^2}\right) \exp\left[-(\bm{p}_{\rho}^2+\bm{p}_{\lambda}^2)b^2\right],
\end{aligned}                              
\label{triton(he3)-wigner function}
\end{equation}                       
where $\bm{\lambda} = \frac{1}{\sqrt{6}}(\bm{r_1}+\bm{r_2}-2\bm{r_3})$ and $\bm{p}_{\lambda} = \frac{1}{\sqrt{6}}(\bm{p_1}+\bm{p_2}-2\bm{p_3})$ are the additional relative coordinate and momentum, respectively. The parameter `b' which is related to their root-mean-square radius is determined to be 2.26 fm~\cite{Ropke:2008}, 1.61 fm and 1.74 fm~\cite{Chen:2003prc} for deuteron, triton and $^3$He, respectively. The phase-space information on protons and neutrons is obtained from UrQMD model. After a Lorentz transformation to the center-of-mass frame of p-n or p-n-n pair, the particles emitted earlier after the last scattering are propagated to the kinetic-out time of other nucleons and a Monte Carlo sampling is used to determine whether the cluster form a nucleus. However, when the collision energy is within the range we are studying, which is \(2.0\sim7.7\) GeV, the yield of light nuclei is not negligible as compared to the proton-neutron multiplicity and proton-neutron pairs repeatedly sampled will influence the final yield of light nuclei. To ensure baryon number conservation, when there are duplicated sampled particles, we determine the generation of cluster based on the size of the Wigner density and exclude the information of generating light nuclei from the phase space distribution of final-state particles. One should notice the sequence of the light nuclei generated by Monte Carlo sampling will affect the final results as shown in Sec.~\ref{resultsA1}.

\begin{figure}[htb]
\setlength{\abovecaptionskip}{0pt}
\setlength{\belowcaptionskip}{8pt}
\centering\includegraphics[trim=1cm 0cm 0.1cm 0.1cm, clip, width=0.48\textwidth]{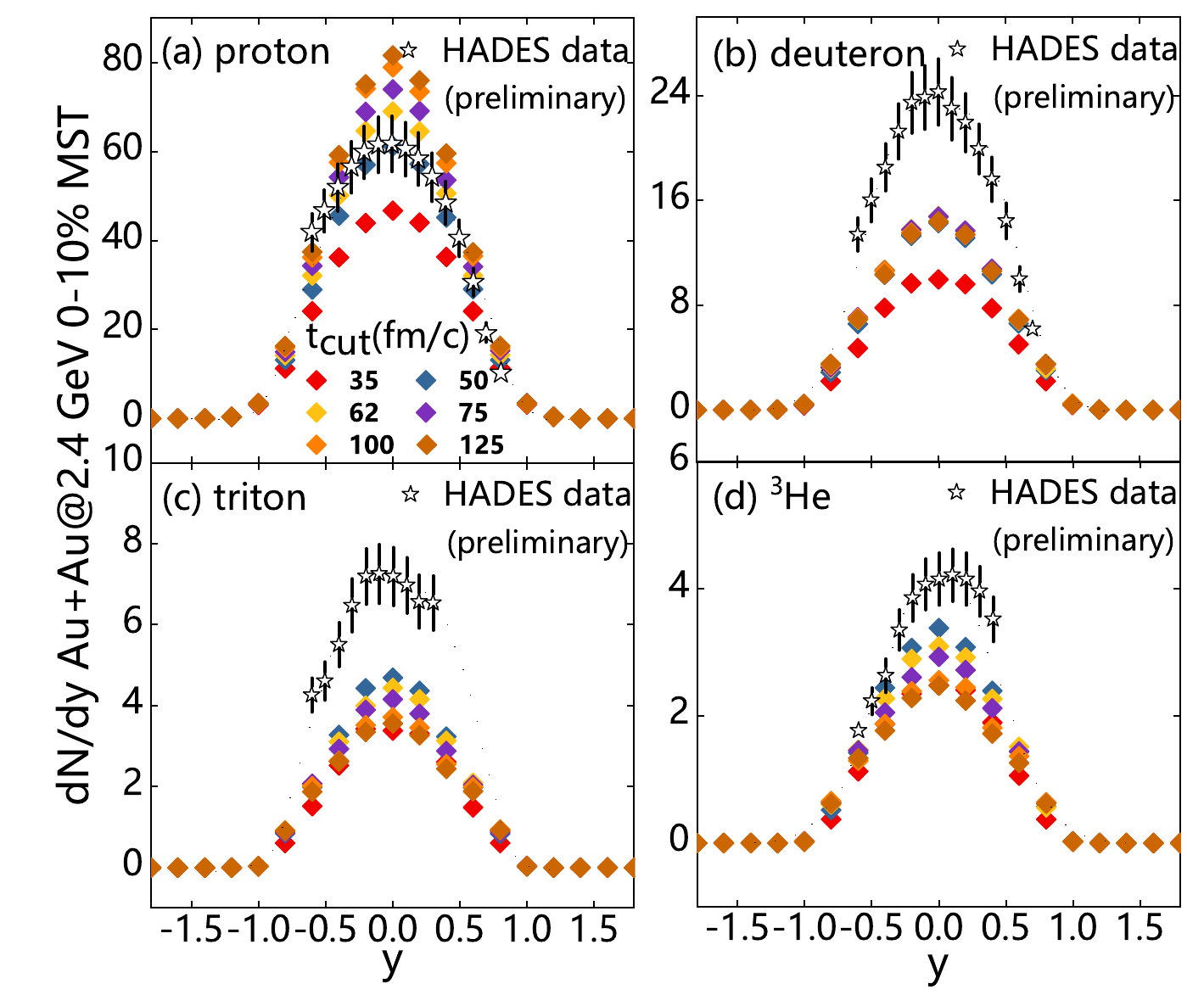}
\caption{Rapidity distribution for: (a) proton, (b) deuteron, (c) triton, and (d) \(^3\)He. Results are from the MST model with various time cuts (35, 50, 75, 100, 125 fm/\(c\)) in 0-10\% Au+Au collisions at \(\sqrt{s_{NN}}\) = 2.4 GeV, comparing with the preliminary HADES data~\cite{Mszala:2019}.}
\label{fig:fig1}
\end{figure}

The Minimum Spanning Tree (MST) method is an algorithm developed to define the nucleon yield in transport models that propagate hadrons~\cite{Aichelin:1991}. This method utilizes the coordinate and momentum information of final-state hadrons with the same timestamp at the moment when the simulation stops. In the MST model we used, a nucleon is considered to belong to a cluster if its spatial and momentum differences between any nucleons in the cluster at their local rest frame satisfy the condition: ${\Delta}r<{\Delta}r_{max},{\Delta}p<{\Delta}p_{max}$, where ${\Delta}r = 3.575\ {\rm fm},{\Delta}p = 0.285\ {\rm GeV}/c$ are adapted in Ref.~\cite{Sombun:2018}. We further explored the ramifications of varying ${\Delta}r$ and ${\Delta}p$ values in our works, which yielded only minor impacts. A notable distinction between the MST and the coalescence model is that MST can form clusters at various times in the final state, typically later than coalescence model. This temporal difference is a factor contributing to the disparity in results between MST and the coalescence model, as discussed in Sec.~\ref{results}. Due to the sequential decay of highly-excited heavier fragments~\cite{Li:2016}, variations in the time cut will alter the multiplicity of light nuclei. Nonetheless, this does not change the rapidity distribution of the light nuclei as displayed in Fig.~\ref{fig:fig1}. 
By comparing outcomes from diverse time cuts, we ascertain that a 50 \(\rm{fm}/c\) duration appears relatively optimal for the production of light nuclei using MST within the energy range we are examining.     

\subsection{HBT correlation function}
\label{HBT model}

To calculate the HBT correlation functions, we use the Lednick\'{y}-Lyuboshitz model~\cite{Lednicky:1981} and the correlation after burner (Crab)~\cite{Pratt:1994} based on the phase space distribution of light nuclei. Both models include the interactions between the final state particles and can be used independently and cross-verified. Lednick\'{y}-Lyuboshitz (LL) technique is a parameterized form used to describe the two-particle correlation function~\cite{Lednicky:2005,Lednicky:2008}. Based on the quantum statistics (QS) and final-state interactions (FSI) which are considered in the LL model, the correlation function of particles emitted with relatively small momentum is sensitive to the information on phase-space and time of the particle source~\cite{Koonin:1977}. In the LL model, the final state interactions of the particles are independent to the particle production process~\cite{Lednicky:1995}, the correlation function of two particles is:
\begin{equation}
\begin{split}
C(\bm{k}^*) = \frac{\int{S(\bm{r}^*,\bm{k}^*)|\psi_{\bm{k}^*}(\bm{r}^*)|^2d^4\bm{r}^*}}{\int{S(\bm{r}^*,\bm{k}^*)d^4\bm{r}^*}},
\end{split}                              
\end{equation}
where \( \bm{r}^* = \bm{x_1} - \bm{x_2} \) is the relative coordinate and \( \bm{k}^* = \frac{1}{2}(\bm{p_1} - \bm{p_2}) \) is half the relative momentum of two particles in their pair rest frame (PRF), \( S(\bm{r}^*, \bm{k}^*) \) is the source emitting function with given \( \bm{r}^* \) and \( \bm{k}^* \). And \(\psi_{\bm{k}^*}(\bm{r}^*)\) is the Bethe-Salpeter amplitude representing the spectrum of the two particles state. When the relative momentum in the PRF is smaller than the inverse Bohr radius, the Coulomb interactions of charged particles will significantly affect the correlation function, enhancing or suppressing the correlation between two particles with like or unlike charges such as p-p, d-d, p-d, t-t and d-t, while for particle pairs containing neutral particles, such as p-n and n-n, only short-range interactions dominated by $s$-wave interaction are involved. The strong interaction is also considered for p-p, p-d and d-d pairs, but not for p-t, d-t and t-t pairs in the LL code. More information about the LL model can be found in Ref.~\cite{Wang:2019,Wang:2023}.

The Crab model~\cite{Crab}, which reads particles' space-time coordinates and four-momentum information at their freeze-out state, integrates quantum statistical effects with Coulomb and strong interactions to calculate the correlation function between particles. By using this model, we can compare the computed output results with experimental data. The correlation functions in Crab is defined as:
\begin{equation}
\begin{split}
C(\bm{P},\bm{q}) = \frac{\int{d^4x_1}\int{d^4x_2}S(x_1,\bm{P}/2)S(x_2,\bm{P}/2)|\phi(\bm{q},\bm{r})|^2}{\int{d^4x_1}S(x_1,\bm{P}/2)\int{d^4x_2}S(x_2,\bm{P}/2)}.
\end{split}                              
\end{equation}
The numerator and denominator are the probability of two particles emitted in the same event and the probability of two single particles emitted in different events. Here, \(\bm{P} = \bm{p_1}+\bm{p_2}\) and \(\bm{q} = \frac{1}{2}(\bm{p_1}-\bm{p_2})\) are the sum and half the relative momentum of the particle pair, \(\bm{r}\) is the relative coordinate. \(S(x,\bm{P}/2)\) represents the probability of emitting a particle with momentum \(\bm{P}/2\) at space point \(x\), \(\phi(\bm{q},\bm{r})\) is the relative wave function of particle pair. Compared to the LL model, Crab only provides final state interaction potentials for a few light clusters, such as p-p and d-d.

\section{Results and discussion}
\label{results}

\subsection{Production of light nuclei}
\label{resultsA1}

\begin{figure*}[htb]
\setlength{\abovecaptionskip}{0pt}
\setlength{\belowcaptionskip}{8pt}
\centering\includegraphics[scale=0.64]{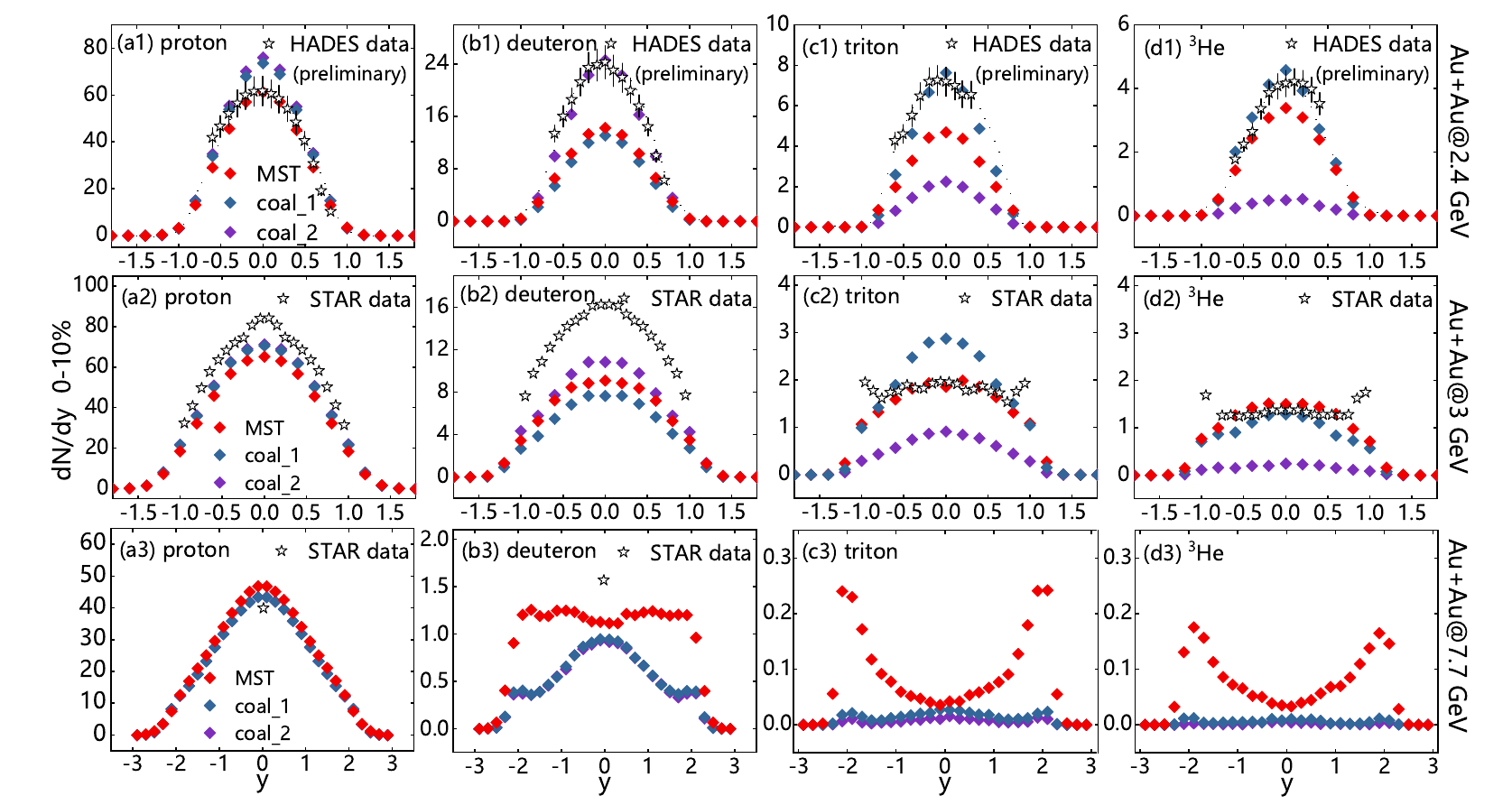}
\caption{Rapidity distribution (dN/dy) for: (a) proton, (b) deuteron, (c) triton, and (d) \(^3\)He. Results are from the MST model (red dots) and coalescence model (blue dots represent the coal\_1 and purple dots represent coal\_2) in 0-10\% Au+Au collisions at \(\sqrt{s_{NN}}\) = 2.4 GeV (upper), 3 GeV (middle), 7.7 GeV (lower) compared with the preliminary HADES data~\cite{Mszala:2019} and the STAR data~\cite{Kozhevnikova:2023mnw,Zhang:2019}.}
\label{fig:fig2}
\end{figure*}

First, we discuss the results of light nucleus production in the center-of-mass collision energy $\sqrt{s_{NN}}$ range of \(2.0\sim7.7\) GeV using the UrQMD combined with MST and coalescence models, and compare them with selected experimental data at 2.4 GeV, 3.0 GeV, and 7.7 GeV~\cite{Mszala:2019,Kozhevnikova:2023mnw,Zhang:2019}. Fig.~\ref{fig:fig2} shows the rapidity distribution of light nuclei produced in Au + Au collisions with a centrality of 0-10\% simulated using both methods. In order to eliminate the influence of ``spectator particles" in the non-collision region, a cut of transverse momentum \(\rm{P}_{\rm{T}}>0.3\) is applied to the final-state particles generated by UrQMD before sampling using the Wigner probability density function.  Meanwhile, in order to ensure baryon number conservation as mentioned in the Sec.~\ref{UrQMD and Light nuclei}, the phase space information of the proton-neutron pairs merged into light nuclei will be removed, then the sequence of the light nuclei generated by Monte Carlo sampling will affect the final result. After reaching chemical equilibrium, the distribution of hadrons within the reaction heat source becomes more uniform. In contrast, at lower collision energies, the increased presence of light fragments diminishes the uniformity of the system distribution. The process of deriving phase space information for light fragment production through coalescence profoundly influences the distribution of proton-neutron wave functions. Consequently, this leads to alterations in their Wigner density, as illustrated with deuterons~\cite{Nagle:1996}:
\begin{equation}
\begin{split}
\frac{d^3N_D}{dP^3_D} = T_r[\hat{\rho}_D\hat{\rho}_{PN}(t)].
\end{split}
\end{equation}
This implies that when there is a high yield of light fragments, the later sampling of light nuclei in coalescence models could introduce a deviation from the actual scenario. Two coalescence models with different sampling orders have been shown in Fig.~\ref{fig:fig2}. The blue diamond represents the light nuclei generation sequence of ``$^{3}$He-triton-deuteron" (coal\_1), while the purple diamond represents the reverse order (coal\_2), and the red diamond represents the results of MST. At the energy point of 2.4 GeV, the description of coal\_1 for the yields of triton and $^{3}$He is better than the results of MST and coal\_2, and the result of the deuteron produced by coal\_1 is lower than coal\_2 and similar to MST due to the conservation of baryon number. It should be noted that due to specific aspects of our coalescence model, the results slightly differ from those in reference~\cite{OmanaKuttan:2023cno}. Utilizing two coalescence methods with distinct sampling sequences can effectively account for the yields of initially sampled light nuclei. However, simultaneously capturing the yields of the entire spectrum of light nuclei remains a challenge. Owing to intrinsic model constraints, the MST method gives lower values for the three light nuclei at 2.4 GeV than experimental data. Nonetheless, based on the results at three energy points, the findings remain notably relevant for discussion. It is noticed that the disparities in coalescence results due to baryon number conservation diminish substantially as the collision energy rises. At 7.7 GeV, for instance, the distinction between the two coalescence results nearly vanishes. At high collision energies, the yield of light nuclei is significantly less than that of protons, and the impact of extracting light fragments becomes negligible on the particle yields. It can be seen that at 7.7 GeV, MST yields more light nuclei than the coalescence model, especially in the higher rapidity region. This is the area where light nuclei could potentially form through the combination of ``spectator particles" and particles from the reaction source. At higher collision energy, a hotter source is forming, making it challenging to form light clusters near the reaction center where rapidity y = 0. Conversely, due to inherent limitations in the model and the extended scattering duration in MST compared to coalescence, MST might generate a greater number of ``spectator light clusters", which may not exist in reality, than the coalescence model. Such discrepancies also influence the outcomes of the correlation function, as discussed in Sec.~\ref{resultsB1}, Sec.~\ref{resultsB2}.

\subsection{Comparison of HBT correlation function calculated by different models with experimental data}
\label{resultsB1}

\begin{figure}[htb]
\setlength{\abovecaptionskip}{0pt}
\setlength{\belowcaptionskip}{8pt}
\centering
\includegraphics[trim=1cm 0cm 0.1cm 0.1cm, clip, width=0.48\textwidth]{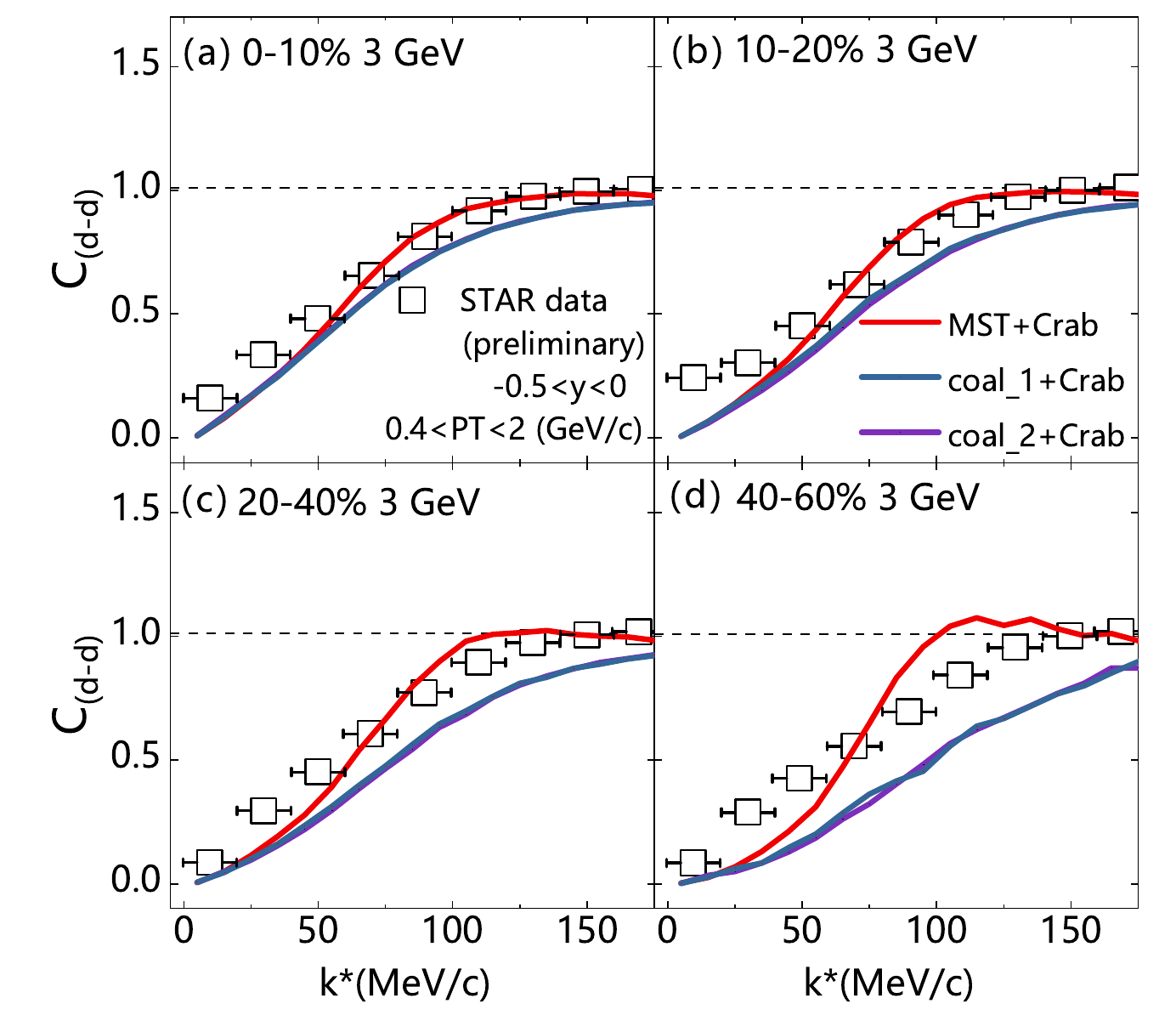}
\caption{D-d correlation functions at midrapidity with four centrality classes in Au + Au collisions at \(\sqrt{s_{NN}}\)=3 GeV compared to the preliminary STAR experiment values~\cite{Mi:2022}. The color lines represent the d-d correlation obtained with the deuteron from MST (red) and coalescence (blue and purple) model, respectively. }
\label{fig:fig3}
\end{figure}

Different mechanisms of light-nucleus production (UrQMD+MST and UrQMD+coalescence) affect not only the yields of the final-state particles, but also their correlation functions. In this section, we analyze the differences in proton-proton and deuteron-deuteron correlation functions derived from two distinct approaches across varying centralities and compare them with the correlation functions measured in experiments. Fig.~\ref{fig:fig3} shows the results of d-d correlation functions at midrapidity with four centrality classes in Au + Au collisions at \(\sqrt{s_{NN}}\)=3.0 GeV compared to the {preliminary} STAR experimental data~\cite{Mi:2022}. The d-d correlation function is calculated by the Crab model and the potentials of d-d pairs are taken from Ref.~\cite{Jennings:1985}. In the low relative momentum region which corresponds to the particle pairs with a closer relative momentum in the reaction heat source, due to the effect of Coulomb repulsion (which dominates the influence of the correlation function), the correlation function of d-d is less than `1'. Although the strong interaction force, mainly contributed to $s$-wave scattering with total d-d spin factor \(S=0,1,2\) have been considered in calculation, the source radius exceeds the effective range of $s$-wave interaction. Consequently, compared to the Coulomb interaction, both the theoretical calculations and the experimental results indicate that the impact of strong interaction is negligible~\cite{Mrowczynski:2021} in d-d correlation functions at this energy point. Since the Coulomb repulsion is anti-correlated with the relative momentum of the pair, the correlation function increases with the pair momentum until it approaches to `1'. It is found although the coalescence methods with different light-nuclei formation sequences produce different light-nuclei yields, the HBT correlation functions of light-nuclei do not show significant differences. The d-d correlation results at 0-40\% centrality generated by the MST approach seem to be in better agreement with the experimental data than coalescence results, while the MST calculation results show a slight increase in the correlation around \(k^*=50\sim150\ {\rm MeV}/c\) region, which is less consistent with the experimental data for the centrality of 40-60\%. The observed situation can be attributed to the involvement of ``spectator particles" in the light cluster formation within the MST algorithm. And because the two light-nuclei algorithms calculate particles at different evolutionary times, as discussed in Sec.~\ref{methods}, the MST method employs the final-state information while the coalescence method uses the phase space information at kinetic freeze-out stage. This leads to a marginally longer expansion period for particles in the MST method and slight differences observed between the correlation functions of the two models. For the d-d correlation functions with coal\_1 + Crab and coal\_2 + Crab, they can reproduce the experimental data at centrality of 0-10\%. As centrality increases, however, the d-d correlation function deviates more from the experimental value. {This leads us to speculate} that in central heavy-ion collisions, a more {  equilibrated} and ideal emitting source {may} be formed. But with the increasing of off-center, the source could be less {equilibrated}, then it makes the correlation function far from the experimental data.

\begin{figure}[htb]
\setlength{\abovecaptionskip}{0pt}
\setlength{\belowcaptionskip}{8pt}
\centering
\includegraphics[trim=0cm 0cm 0cm 0cm, clip, width=0.48\textwidth]{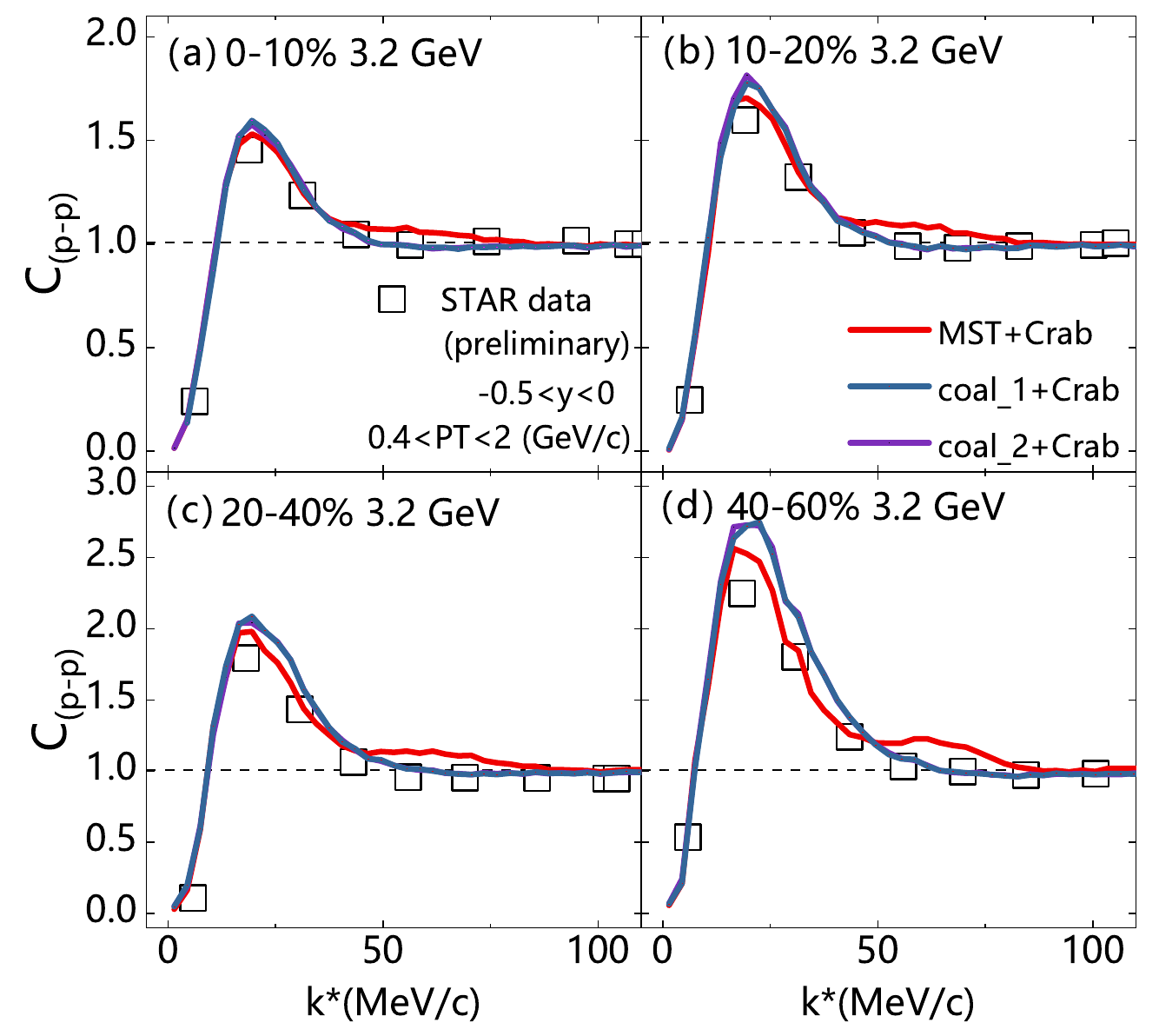}
\caption{P-p correlation functions at midrapidity with four centrality classes in Au + Au collisions at \(\sqrt{s_{NN}}\) = 3.2 
 GeV compared to the {preliminary} STAR experiment value~\cite{QM:2023}. The color lines represent the p-p correlation function obtained with the proton from the MST (red) model and the coalescence (blue and purple) model, respectively.}
\label{fig:fig4}
\end{figure}  

The results of p-p correlation function at midrapidity with four centrality classes compared to the {preliminary} STAR data at \(\sqrt{s_{NN}}\) = 3.2 GeV~\cite{QM:2023} are presented in Fig.~\ref{fig:fig4}. At lower relative momentum, the Coulomb repulsion dominates in the interaction and the $s$-wave function with antisymmetry between two protons together suppress the correlation. With the increasing of relative momentum, due to the combined effect of $s$-wave attraction in the final state strong interaction between protons, the correlation function peaks at the relative momentum \(k^*=20\ {\rm MeV}/c\). The p-p correlation functions calculated by the two models are basically consistent with the experiment data and similarly, when using the coalescence method, the order of light nucleus generation does not affect the correlation function, which indicates that the emission source radius of light clusters generated by the coalescence method is consistent. But one can notice that for large centrality at 40-60\%, the p-p correlation function calculated by the coalescence model is slightly higher than experimental data, which can be explained similarly to the previous case for the d-d correlation. Additionally, we observed a slight increase in the MST results for p-p in the \(k^*=50\sim100\ {\rm MeV}/c\) region. This can be attributed to the same possibilities that the ``spectator particles" may combine with particles from reaction source in the MST model, resulting in a more compact proton-proton source in this relative momentum region. 

\begin{figure}[htb]
\setlength{\abovecaptionskip}{0pt}
\setlength{\belowcaptionskip}{8pt}
\centering
\includegraphics[trim=0cm 0cm 0cm 0cm, clip, width=0.48\textwidth]{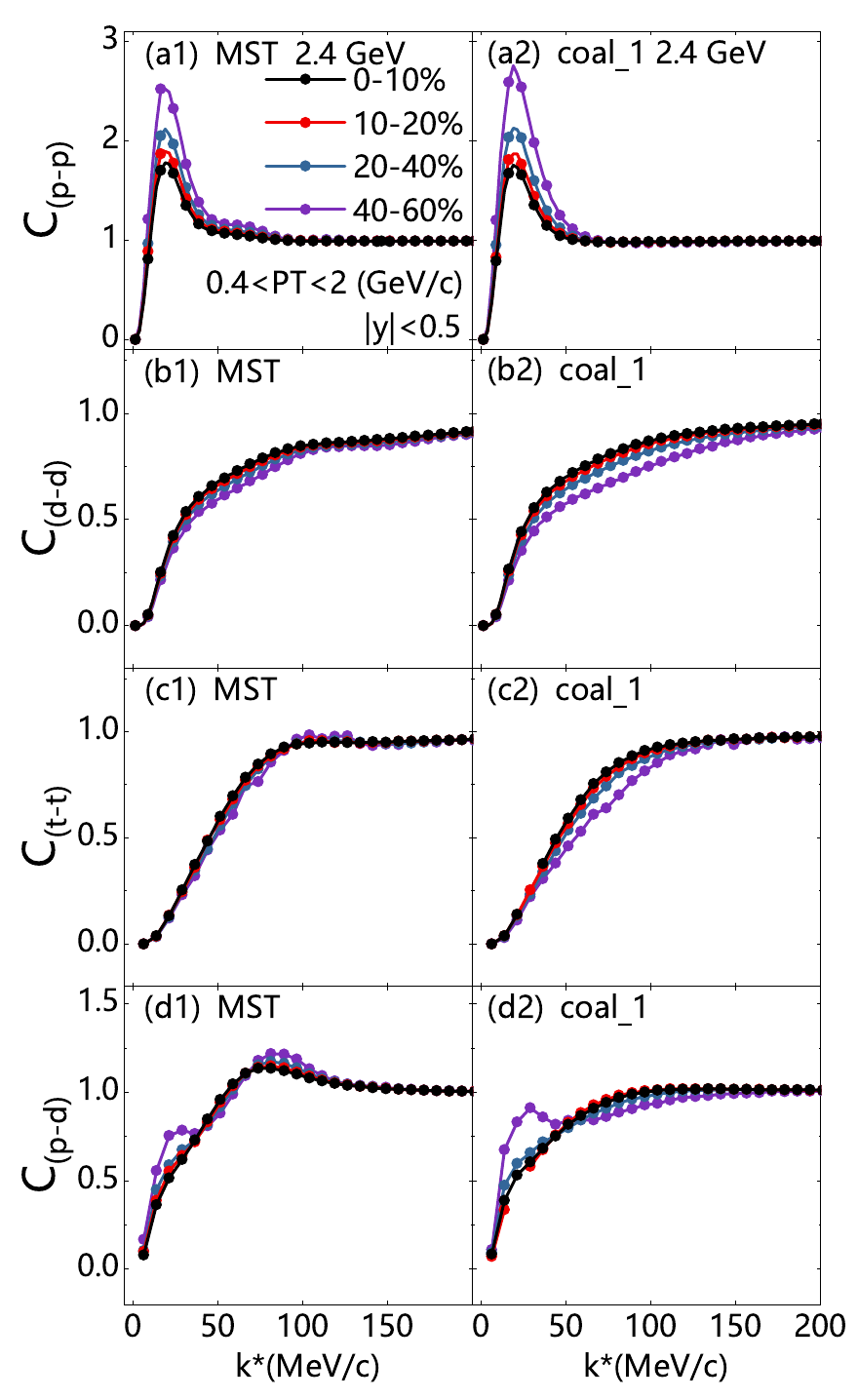}
\caption{Correlation functions of p-p (a), d-d (b), t-t (c), and p-d (d) at midrapidity in four centrality classes for Au + Au collisions at \(\sqrt{s_{NN}}\) = 2.4 GeV. The left column is the result of MST + L-L model, and the right column is coal\_1 + L-L. The color dot-lines represent different collision centralities (black for 0-10\%, red for 10-20\%, blue for 20-40\% and purple for 40-60\%), respectively.}
\label{fig:fig5}
\end{figure}

\begin{figure}[htb]
\setlength{\abovecaptionskip}{0pt}
\setlength{\belowcaptionskip}{8pt}
\centering
\includegraphics[trim=0cm 0.5cm 7cm 0cm, clip, width=0.48\textwidth]{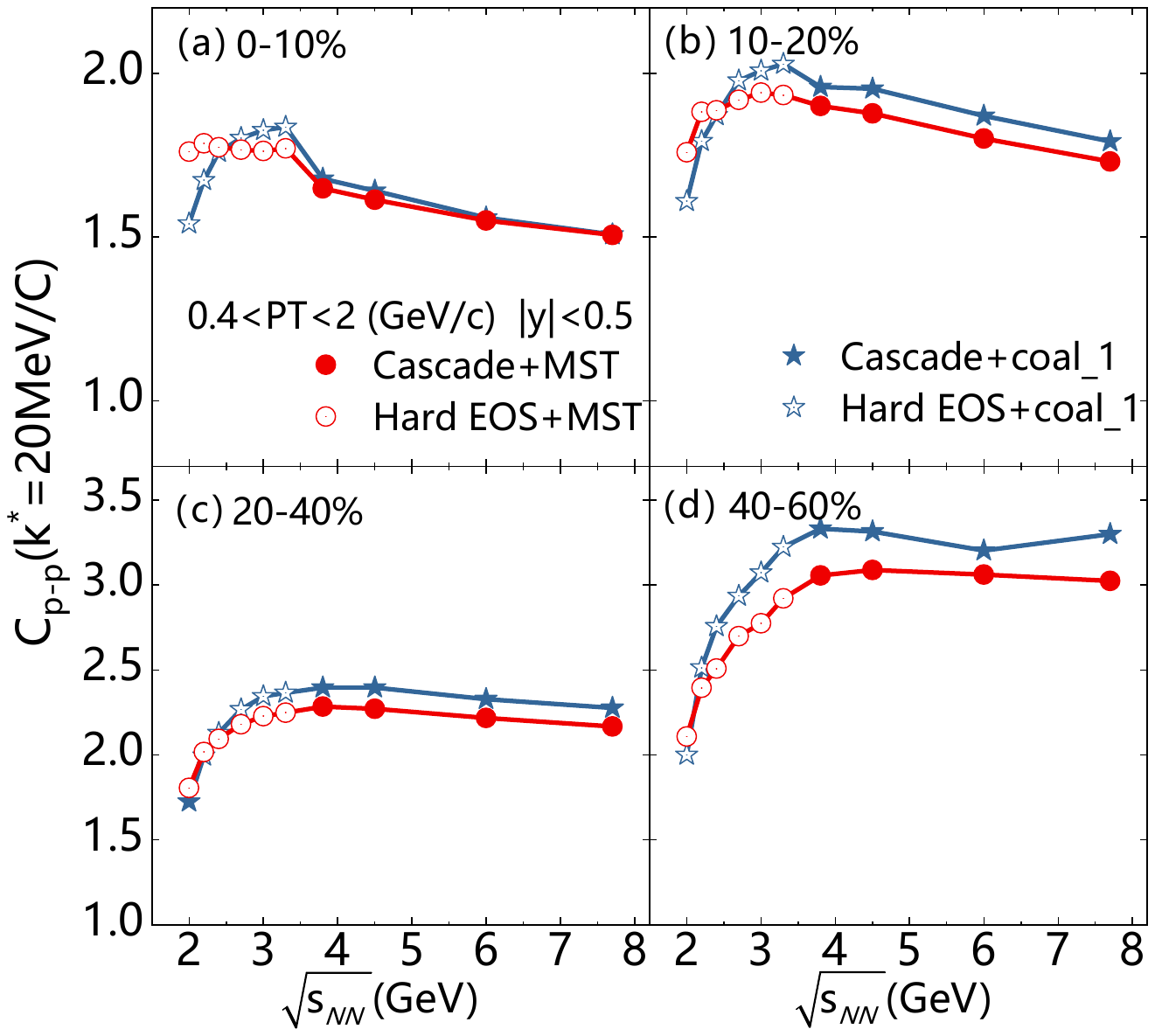}
\caption{The peak of proton-proton correlation (at around 20 \( \rm{MeV}/c \) relative momentum \(k^*\)) at midrapidity varies with collision energy. The red dot-line represents the results of the MST method in four collision centralities, and the blue star-line represents the results by the coalescence method. {Solid symbols and hollow symbols represent the UrQMD model in potential mode and cascade mode, respectively.}}
\label{fig:fig7}
\end{figure}

\begin{figure}[htb]
\setlength{\abovecaptionskip}{0pt}
\setlength{\belowcaptionskip}{8pt}
\centering
\includegraphics[trim=0.1cm 7cm 5cm 0.1cm, clip, width=0.48\textwidth]{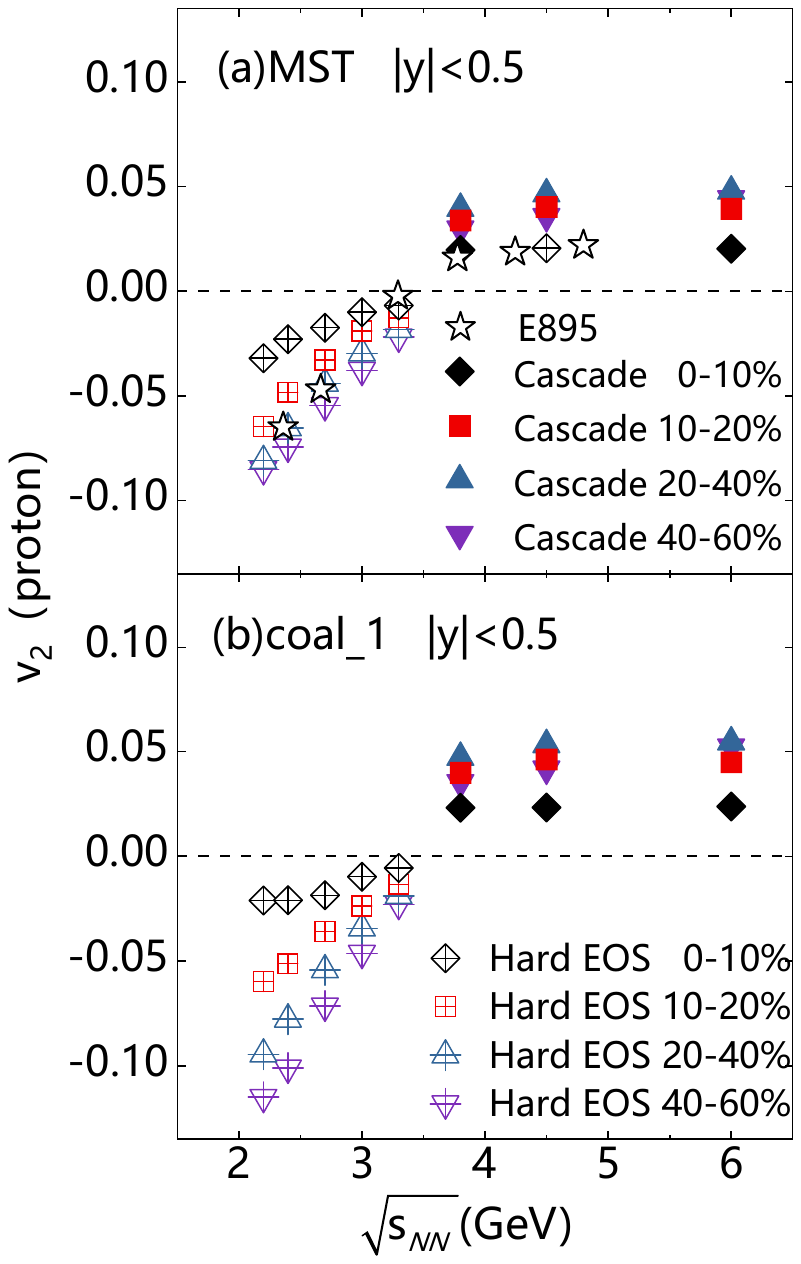}
\caption{The elliptic flow \(v_2\) for proton at midrapidity in four centralities (empty star for experimental data~\cite{E895:1999}, black for 0-10\%, red for 10-20\%, blue for 20-40\%, and purple for 40-60\%) for Au + Au collisions at \(\sqrt{s_{NN}}=2.0\sim7.7\) GeV. {Solid symbols and hollow symbols represent the UrQMD model in potential mode and cascade mode,  respectively.} The upper panel is the results with the MST method and the lower panel  is the results with the coal\_1 method).}
\label{fig:fig8}
\end{figure}

\subsection{HBT correlation of light nuclei}
\label{resultsB2}


By using the L-L model, the correlation functions of light nuclei at midrapidity with four centrality classes (Au + Au at \(\sqrt{s_{NN}}=2.4\ \rm{GeV}\)) are shown in Fig.~\ref{fig:fig5}. We also compare two different light nuclei production models here: the left column is the result of MST, and the right column shows coal\_1 (since different coalescence methods produce the same correlation function as shown in Sec.~\ref{resultsB1}, we choose the coal\_1 to discuss). The correlations of proton-proton and deuteron-deuteron pairs are shown in Fig.~\ref{fig:fig5}(a1), (a2), (b1) and (b2). Both results indicate that the reaction source is smaller in peripheral collisions. Due to the coulomb interaction, which is the only final-state interaction for t-t considered in the L-L code, the t-t correlation functions get low value at low relative momentum and increase with the relative momentum. The t-t production by the MST method is slightly different from the coalescence approach, that the results of the coalescence approach are more sensitive to centrality and the particles in peripheral collision emit from a smaller source. The p-d correlation shows a distinct peak at low relative momentum in peripheral collisions within both models, which could be attributed to $s$-wave attraction as in p-p correlations. In peripheral collisions, where the source size is smaller, the strong interactions between particles appear to impart a noticeable influence on the correlation function, contrary to being negligible in central collisions. Notable differences are also seen in the p-d correlation functions between the two models; specifically, the MST results show a slight peak in the \(k^*=50\sim100\ {\rm MeV}/c\) region. This can be attributed to the same reason as discussed before. 

The light nuclei correlation functions in the midrapidity region in the energy range of \(\sqrt{s_{NN}}=2.0\sim7.7\ \rm{GeV}\) will be discussed below. Fig.~\ref{fig:fig7} shows the peak of proton-proton correlation (at around 20 \( {\rm MeV}/c \) relative momentum \(k^*\)) varies  with collision energy. The correlation results of p-p generated by the MST and coal\_1 methods combined with the L-L model both exhibit non-monotonic behavior in this range. A notable turning point is observed between \(3.0\sim4.0\) GeV. We speculate that this is linked to the collective flow of the final state proton. As suggested in Ref.~\cite{Danielewicz:2002}, at low energies, the expanding particles within the reaction source is squeezed out by the presence of the spectator matter. This suppresses in-plane emission, leading to a shift towards out-of-plane emission. This corresponds to the energy region where \(v_2\), which describes the elliptic flow of the emitted particles, is negative~\cite{FOPI:2004}. As the collision energy increases, spectator matter passes the compressed zone and the particles emitted into the reaction plane. Then, suppression effect disappears, leading to a transition of proton correlation in Fig.~\ref{fig:fig7} and the change of \(v_2\) is from negative to positive. As shown in Fig.~\ref{fig:fig8}, the \(v_2\) of UrQMD+MST and coalescence with different collision centralities both show a transition from negative to positive between \(3.0\sim4.0\) GeV, which is consistent with experimental data~\cite{E895:1999}. As the collision energy increases from 2 GeV, the system creates more particles and they are constrained in a volume due to spectators. Thus particle correlation becomes stronger. With increasing collision energy again and the rise in the elliptic flow \(v_2\), the source size has been much expanded at higher energies since the suppression effect vanishes, resulting in a weakened p-p correlation function with the increasing of source size. Meanwhile, due to the same reason discussed in Sec.~\ref{resultsB1}, the correlation functions of the MST method are smaller than the coal\_1 method at different collision energies. We also explored other light nuclei correlation functions, such as d-d and t-t pairs, as a function of collision energy. However, due to the tiny differences among these correlation functions, they show small energy dependence within this energy range. Specifically, in the 2.0$\sim$3.0 GeV energy range, the light nuclei are emitted from a smaller source as the collision energy increases. A transition occurs between \(3.0\sim4.0\) GeV, after which the reaction source becomes larger as the energy increases. These findings mean that the collective flow of the final particles is related to the momentum correlation function of the particles. The collective motion of the final-state particles is associated with the emission of the source, and the change of collective flow of particles may be reflected in the correlation functions of particles. And as the discussions in Ref.~\cite{PDanielewicz1998}, one can get that the analysis of elliptic flow can provide invaluable constraints for the EOS of nuclear matter. From the correlation here, it implies that the correlation functions of particles could be sensitive to the EOS.

\section{Conclusions}
\label{summary}
We presented the yield of light nuclei in the energy range of ($\sqrt{s_{NN}}$=2.0$\sim$7.7 GeV) for Au + Au collisions by coupling either the coalescence method or the Mini-Spanning-Tree method with the Ultra-relativistic-Quantum-Molecular-Dynamics model. It is found that the increased production of light nuclei in this energy region results in a high $d/p$ or $t/p$ ratio. This indicates that the formation of light clusters at this energy is intricate. Different sequencing in the formation of light nuclei leads to varying outcomes when using the Wigner-density coalescence methods and quantitative differences with the results of the MST method. Upon comparing the yield of light nuclei results with experimental data, we observed that both algorithms could fit the experiment data in part but can cover all light nuclei production due to the defects of the model itself. And we also predict light nuclei correlation functions in this energy region. By comparing the correlation functions with experimental data, we found that although the yields of light nuclei constructed by two methods differ quantitatively, the calculated light nuclei correlation functions in central collisions are close to each other and both can describe the experimental values. While in peripheral collisions, {due to the possibility that the source being less equilibrated} and the MST method may generate clusters combined with ``spectator particles", the correlation functions may differ slightly from the experimental values. We also analyzed the light nuclei correlation functions at midrapidity with different collision centrality and found that light nuclei emission from a smaller source in peripheral collisions. Furthermore, we give the proton-proton correlation function at midrapidity in energy range of \(\sqrt{s_{NN}}=2.0\sim7.7\ \rm{GeV}\). Through analyzing the collective flow of protons, we associate the momentum correlation functions of protons with the elliptic flow, positing that the observed trends in the proton correlation function with energy arise from the squeeze out effect seen in the elliptic flow of protons. We also examined the correlation functions of other light nuclei and observed a consistent trend. These observations indicate that the collective flow of particles is related to their correlation functions. Moreover, changes in the collective flow are likely to be reflected in the variations observed within the  correlation functions of particles. For future study, we expect that  more experimental data of light nuclei in this energy region can be available to compare with our calculations.

\begin{acknowledgments}

Authors thank Drs. Kai-Jia Sun, Song Zhang and Peng-cheng Li for communications.
This work was supported in part by the National Natural Science Foundation of China under contract Nos. 11890710, 11890714, and 12147101, the Strategic Priority Research Program of Chinese Academy of Sciences under Grant No. XDB34000000, the Guangdong Major Project of Basic and Applied Basic Research No. 2020B0301030008, and the National Key R\&D Program of China under Grant No. 2018YFE0104600 and 2016YFE0100900. 

\end{acknowledgments}

\end{document}